\documentclass[prl,twocolumn,showpacs,epsfig]{revtex4}
\usepackage{color,graphicx,bbold}
\usepackage{amsmath,amssymb,amsthm}
\usepackage{epsfig}
\usepackage{epstopdf}
\usepackage{color}
\usepackage{enumerate}
\DeclareMathOperator{\tr}{\rm Tr}

\newcommand{\ket}[1]{\left\vert#1\right\rangle}
\newcommand{\bra}[1]{\left\langle#1\right\vert}
\newcommand{\deter}[1]{\left |#1\right |}
\newcommand{\deterv}[1]{\left |\left |#1\right |\right |}

\begin{document}

\title{Geometrical characterization of non-Markovianity}

\author{Salvatore Lorenzo$^{1,2}$, Francesco Plastina$^{1,2}$, and Mauro Paternostro$^{3}$}
\affiliation{$^1${Dipartimento di  Fisica, Universit\`a della Calabria, 87036 Arcavacata di Rende (CS), Italy}\\
$^2$ INFN - Gruppo collegato di Cosenza \\
$^3$ Centre for Theoretical Atomic, Molecular, and Optical Physics, School of Mathematics and  Physics, Queen's University, Belfast BT7 1NN, United Kingdom}
\date{\today}

\begin{abstract}
We introduce a new tool for the quantitative characterisation of the departure form Markovianity of a given dynamical process. Our tool can be applied to a generic $N$-level system and extended straightforwardly to Gaussian continuous-variable systems. It is linked to the change of the volume of physical states that are dynamically accessible to a system and provides qualitative expectations in agreement with some of the analogous tools proposed so far. We illustrate its prediticve power by tackling a few canonical examples.
\end{abstract}

\maketitle

The interaction with an environment leads a quantum system to
dissipate energy and lose its coherence. The process, however,
needs not be monotonic and the system may temporarily recover some
of the lost energy and/or information. This is the essence of a
non-Markovian behavior, which can be characterized and quantified
in many different ways \cite{Wolf2008b,Chruscinski2010}. One
possibility is to look for temporary increases of the entanglement
shared by the system with an isolated ancilla, which amounts to measure the
deviation from divisibility of the dynamical map describing the
system's reduced evolution~\cite{Huelga2010} (RHP measure). A different
approach \cite{Breuer2009,Breuer2010} relies on measuring the
distinguishability of two {\it optimal} initial states that have evolved
through the same quantum channel, looking for any non-monotonicity (BLP measure). Further proposals have been put forward, based on the decay rates
entering the master equation \cite{andersson2010}, on the Fisher
information flow \cite{sun2010}, on the use of the quantum mutual
information \cite{luo12} or of channel capacities \cite{bogna},
and on spectral considerations \cite{spectra}. This variety of tools highlight the multi-faceted nature of the problem embodied by a characterisation of non-Markovian
dynamics and its inherent difficulty, which prevents the formulation of a unique tool.

In this paper, we contribute to the quest for sharp tools able to capture the various aspects with which non-Markovianity manifests itself and propose a method that qualifies non-Markovian evolutions based on the rate of change of the volume of accessible
states of the evolved system.  As for the divisibility-based
approach \cite{Huelga2010}, this is a characteristic of the map
itself that does not depend on the
initial state(s) of the system (nor needs to be optimized over them). A quantum evolution is Markovian if it is an
element of any one-parameter continuous completely positive
semigroup: in this case the process is unidirectional and there is
no recovery of energy/information/coherence by the system. This
implies that the domain's volume of the dynamical map decreases
monotonically. On the contrary, we associate non-Markovianity of
the dynamics to a growth of this domain's volume. We thus
define a quantifier of non-Markovianity as the sum of the
(temporary) volume increases which occur during the time
evolution.

In the case of a single qubit, this can be linked to the BLP
measure~\cite{Breuer2009}, as the trace distance coincides with the
Euclidean distance on the Bloch sphere
and the pair of states that maximize the measure lie on the
boundary of the convex subspace of physical states \cite{wibeta}.
As a result, if the trace distance decreases monotonically, so does the volume, which is however much easier to
evaluate through the determinant of the dynamical map, as we address in details in this paper. Our aim here is to
formalize the intuition at the basis of our proposal towards the construction of a new tool for the quantitative characterisation of
non-Markovianity. It should be stressed that ours is not {\it yet another attempt at the quantification of the degree of non-Markovianity} of a given dynamical process,
but the proposal for a novel way to reveal effects of an evolution departing from the features of Markovianity that went so far overlooked.  Our proposal enjoys features
of practicality and intuition of interpretation that are somehow missing from analogous, otherwise equally valid quantifiers.

{\it Systems with finite dimensional Hilbert spaces}. Irrespective
of the initial open system state, a reduced time evolution derived
from the unitary dynamics of a larger system can always be
described by a linear, Hermitian map \cite{Lidar2009}, not
necessarily completely positive  due, e.g., to the presence of
initial system-environment correlations \cite{jordan2004}. A
Markovian or memory-less behavior leads to master equations in the
Lindblad form \cite{Lindblad1976,Gorini1976}, with the map obeying
the semigroup composition law.  We consider a positive trace
preserving map
\begin{equation}\label{map}
\phi_t:\hat{\rho}(0)\rightarrow\hat{\rho}(t)=\phi_t[\hat{\rho}(0)]
\end{equation}
for the quantum state of a $N$-level open system, which can be
expressed through a generalized Bloch vector $\mathbf{r}$, whose
components are the expectation values of the traceless, hermitian
generators of SU($N$), $G_i (i = 1, . . . ,N^2-1)$, for which
$\tr[\hat{G}_i \hat{G}_j]=\delta_{ij}$. By including the identity
$\hat{G}_0=\hat{\mathbb{I}}/\sqrt{N}$, any state can be written as
\begin{equation}
\hat{\rho}=\sum_{\alpha=0}^{N^2-1} \tr[\hat{\rho} \hat G_{\alpha}] \, \hat G_{\alpha}
\equiv \sum_{\alpha=0}^{N^2-1}r_\alpha\hat{G}_\alpha
\end{equation}
with $\vec{r} = (1/\sqrt{N}, \mathbf{r})$. A systematic construction of
the $\{ \hat{G}_\alpha \}$ is given in Refs.~\cite{Mahler,Alicki} and leads us to $\lbrace \hat{G}_\alpha \rbrace_{\alpha=1}^{N^2-1} = \lbrace \hat{u}_{jk}, \hat{v}_{jk}, \hat{w}_{l}  \rbrace /\sqrt{2}$ with
\begin{equation}
\begin{aligned}
&\hat{u}_{jk} = \ket{j}\bra{k} + \ket{k}\bra{j}\, ,\;\;\;\;\;\;\;\hat{v}_{jk} = -i(\ket{j}\bra{k} - \ket{k}\bra{j}),\\
&\hat{w}_{l}  = \sqrt{\frac{2}{l(l+1)}}\sum\nolimits_{j=1}^{l} (\ket{j}\bra{j} - l\ket{l+1}\bra{l+1}),
\label{SUNgen}
\end{aligned}
\end{equation}
where the span of the indices is such that $1 \le j < k \le N,1 \le l \le N-1$ and $\{ \ket{m} \}^{N}_{m=1}$ is an orthonormal basis of the
open system's Hilbert space. This gives Pauli spin operators for
$N=2$ and Gell-Mann
operators for $N=3$.

Writing the map in Eq.~\eqref{map} in this basis, one gets
\begin{equation}
\vec r_t = \mathbf{F}(t) \, \vec r_0 \, , \quad \mbox{with} \;
F_{\alpha\beta}(t) =\tr[\hat{G_\alpha}\phi_t[\hat{G}_\beta]]
.\label{mapmatrix}\end{equation}
%
As $F_{0\beta}(t)=\delta_{0\beta}$, this is an affine
transformation for the Bloch vector. Letting $q_\beta(t)=F_{\beta
0}$, we have
\begin{equation}
\mathbf{F}(t) = \left(\begin{matrix}
1 & \mathbf{0}\\
\mathbf{q}_t & \mathbf{A}_t
\end{matrix}\right)
\rightarrow
\mathbf{r}_t=\mathbf{A}_t\mathbf{r}_0+\mathbf{q}_t/\sqrt{N}.
\label{affine_map}\end{equation}
The real matrix $\mathbf{A}_t$ can be decomposed as $
\mathbf{A}_t=\mathcal{O}^1_t \mathbf{D}_t {\mathcal{O}_t^2}^T $,
where $\mathcal{O}^n_t$'s are orthogonal matrices and $\bf D$
is a positive semi-definite diagonal one. In what follows, we will indicate with
$\deter{M}$ the determinant of a matrix $M$.
The findings above imply that
$\deter{\mathbf{F}_t}=\deter{\mathbf{A}_t}=\deter{\mathbf{D}_t}$.
The action of $\mathbf{F}$ is given by a first rotation
(possibly composed with an inversion), then a shrink of the Bloch
vector followed by a final rotation plus a translation.
Its determinant gives the contraction factor for the volume of
accessible states, given by the measure of the set of evolved
Bloch vectors, with respect to its value at $t=0$.

The set of physical Bloch-vectors for an $N$-level system is given
by \cite{Kimura2003}
\begin{equation*}
B_N = \{ \mathbf{r} \in {\mathbb{R}}^{N^2-1} :(-1)^j
a_j({\mathbf{r}}) \ge 0 \ (j=1,\ldots,N)\},
\end{equation*}
where $a_j(\mathbf{r})$ are the coefficients of the characteristic
polynomial $\det(x\mathbb{I}^N-\hat{\rho})$ with
$\hat{\rho}=\frac{1}{N}\mathbb{I}^{(N)}+\sum_{i=1}^{N^2-1}r_i \hat
G_i$. In spherical coordinates, the volume element of $B_N$ is
\begin{equation}
d^NV =\left |\deter{\frac{\partial
(r_i)}{\partial(R,\phi_j)}}\right |dR\,d\phi_1 \, d\phi_2\cdots
d\phi_{n-1} \,
\end{equation}
and, by direct substitution, it is straightforward to check that any positive trace-preserving map described by Eq.~\eqref{mapmatrix} induces the change
\begin{equation}
d^NV(t) = \left |\deter{\mathbf{A}_t}\right | d^NV(0).
\end{equation}
%
%
Therefore, $\left|\deter{\mathbf{A}_t}\right|$ describes the change in
volume of the set of states accessible through the evolution of the reduced state. In particular, $\left|\deter{\mathbf{A}_t}\right|$
decreases monotonically for any positive, linear and trace
preserving map~\cite{Wolf2008a}, and so it does
for any element of a completely positive continuous one-parameter
semigroup. Indeed, if $\phi_t=e^{t \mathcal{L}}$ with
\begin{equation}
\mathcal{L} \, \hat{\rho} = i [\hat{\rho},H] + \sum_{\alpha,\beta}
\gamma_{\alpha,\beta} \left(C_\alpha \hat{\rho} C_\beta^\dagger
-\frac12\{C_\beta^\dagger C_\alpha,
\hat{\rho}\}\right)\label{LinForm}
\end{equation}
$\gamma\geq 0$, and $H$ the Hamiltonian of the system, we get $|
\mathbf{A}_t|=e^{-N\tr[\gamma]}$, which is a a constant.

\begin{figure}[t]
\centering{\bf (a)}\hskip4cm{\bf (b)}\\
\includegraphics[width=8.7cm]{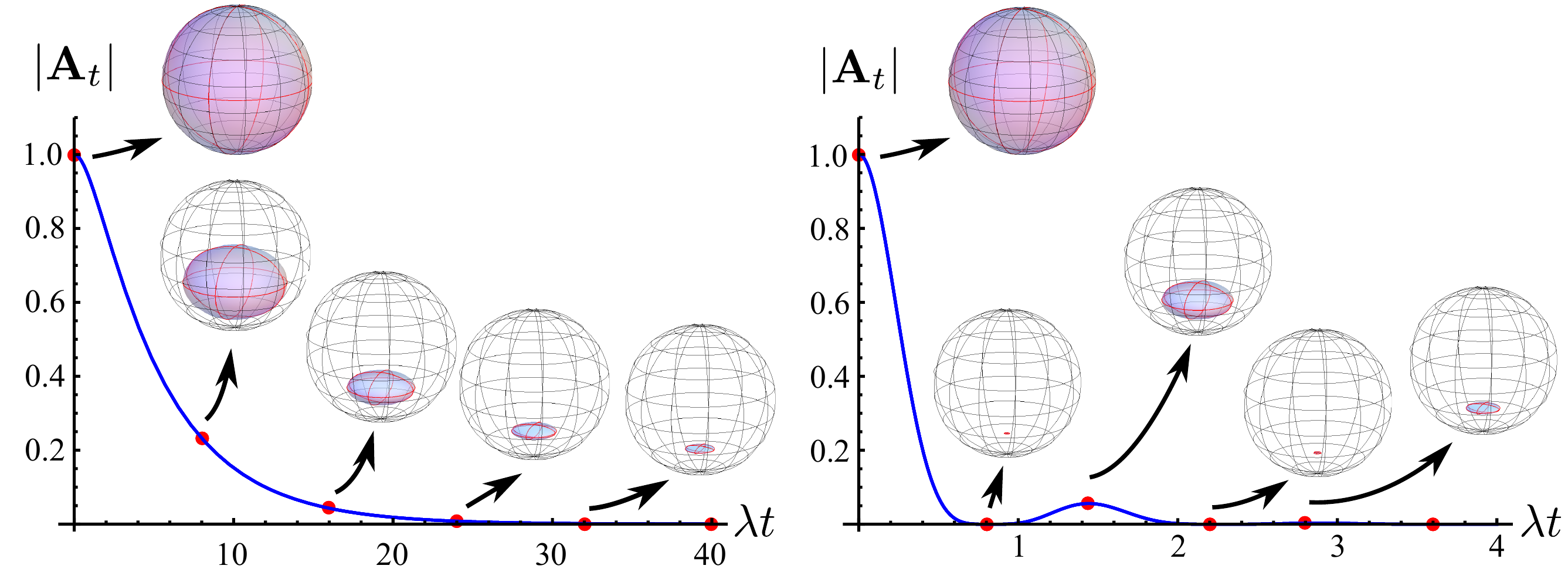}
\caption{(Color online)  Time evolution of the determinant of the
map for a generic Markovian [panel {\bf (a)}] and non-Markovian dynamics [panel {\bf (b)}].
Quantitatively, the curves displayed in the figure correspond to the spontaneous emission of a two-level system in a
resonant leaky cavity (cf. {\it Example 1}) with the Lorentzian spectral density given in Eq.~(\ref{spectraldens}). Panel {\bf (a)}
shows the (Markovian) case with $\gamma_0/\lambda=0.1$ (bad cavity
limit), while panel {\bf (b)} shows a (non-Markovian) evolution in
the good cavity limit with $\gamma_0/\lambda=10$. We also picture the set of accessible states, whose volume changes in time
according to the behavior of the determinant. } \label{mnm}
\end{figure}

Time dependent generators of the form in Eq.~\eqref{LinForm}
with $\gamma(t)\geq0$ lead to time-dependent Markovian
processes.
The dynamical map $\phi_{t+\tau,t}= \exp
[T\int_{t}^{t+\tau}\mathcal{L}dt]$, although not being part of a
dynamical semigroup, is divisible and can be written as the
composition of two CPT maps
\begin{equation}
\phi_{t+\tau,0}=\phi_{t+\tau,t}\;\phi_{t,0}~~(\forall \tau,t\geq 0).
\label{divisibility}\end{equation} As a consequence, in this case
too the determinant is monotonically decreasing. These
considerations lead us to define a new way to quantify the
non-Markovian character of a quantum evolution through the
variation of the volume of accessible states
\begin{equation}
\mathcal{N}_V=\frac{1}{V(0)}\int_{\frac{dV(t)}{dt}>0}\frac{dV(t)}{dt}=\int_{\frac{d\left|\deter{\mathbf{F}_t}\right|}{dt}>0}\frac{d\left|\deter{\mathbf{F}_t}\right|}{dt}
\label{NM_nostra}
\end{equation}
The intuitive meaning of this definition is illustrated in Fig.
\ref{mnm}, where the time evolution of the determinant is
explicitly shown for the case of a two-level atom spontaneously
decaying in a structured environment (cf. {\it Example 1} for details).
The monotonous volume decay characterizing a Markovian time
evolution is contrasted with a non-Markovian behavior in which a non-zero
accessible volume re-appear after being fully decayed.

Besides the geometric  interpretation, such a measure has a simple
physical meaning based on the change of the classical information
encoded in the states. Suppose that a set of quantum states is
given, whose elements are characterized by an arbitrary
distribution of Bloch vectors within $B_N$, described by a
probability density $p(\mathbf{r})$. The corresponding differential entropy is
\begin{equation}
  h[p(\mathbf{r})]=-\int_{B_N} p(\mathbf{r})\, \log_2p(\mathbf{r}) \, d V_N.
\end{equation}
If such states are taken as the initializations of the map
$\phi_t$, after a time $t$ the probability density function
is re-scaled as $p'(\mathbf{r}_t)= p(\mathbf{r}_t)/\deterv{A}$ and
the entropy changes accordingly as
\begin{equation}
h[p'(\mathbf{r}_t)] {-} h[p(\mathbf{r}_0)] {=} log_2
\deterv{\mathbf{A}_t}.
\end{equation}
Thus, a contraction of volume is equivalent to a loss of classical
information.

The BLP measure also enjoys an information theoretical interpretation,
but a comparison between the two quantifiers is difficult. Indeed, although it
is known that the optimal states that enter the measure proposed in Ref.~\cite{Breuer2009} lie on the boundary of the volume accessible throughout the dynamics~\cite{wibeta},
this does not necessarily imply a connection with the measure of
such volume. On the other hand, the difference between ${\cal
N}_{V}$ and the RHP measure is simpler to describe: as the
determinant is contractive under composition of positive maps, it
follows that it does not increase whenever the intermediate map
$\phi_{t+\tau,t}$ in Eq.~(\ref{divisibility}) is non positive. The
entanglement-based measure of Ref.~\cite{Huelga2010}, on the other hand, is
non-zero in the less restrictive condition of the map being
not completely positive. Therefore, if $\phi_{t+\tau,t}$ is
positive but not completely positive, the map is non-divisible. Nonetheless, we find
${\cal N}_{V}=0$.

\begin{figure}[hbtp]
\centering {\bf (a)}\hskip3.5cm{\bf (b)}\\
\includegraphics[width=3.5cm]{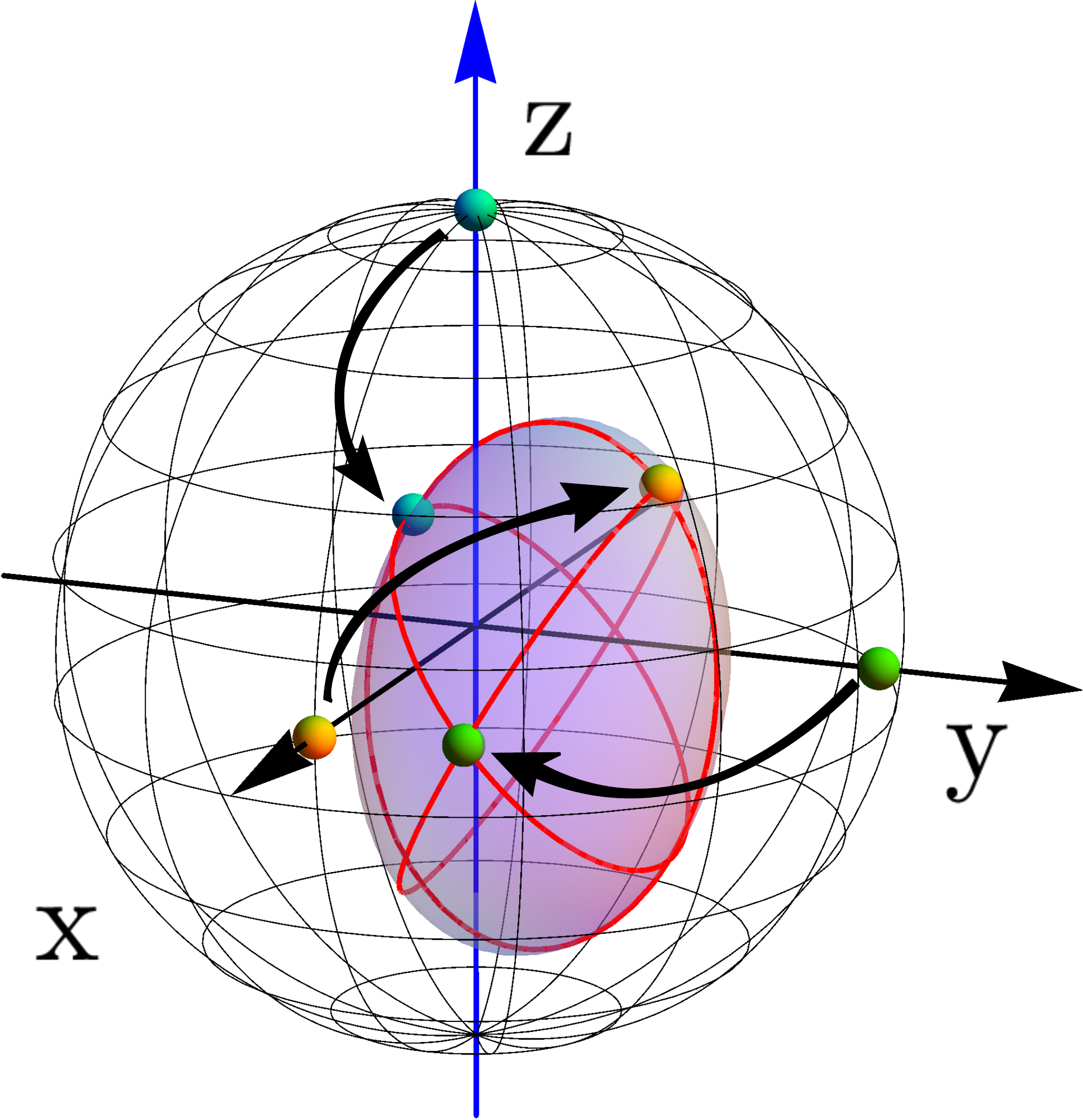}
\includegraphics[width=4.5cm]{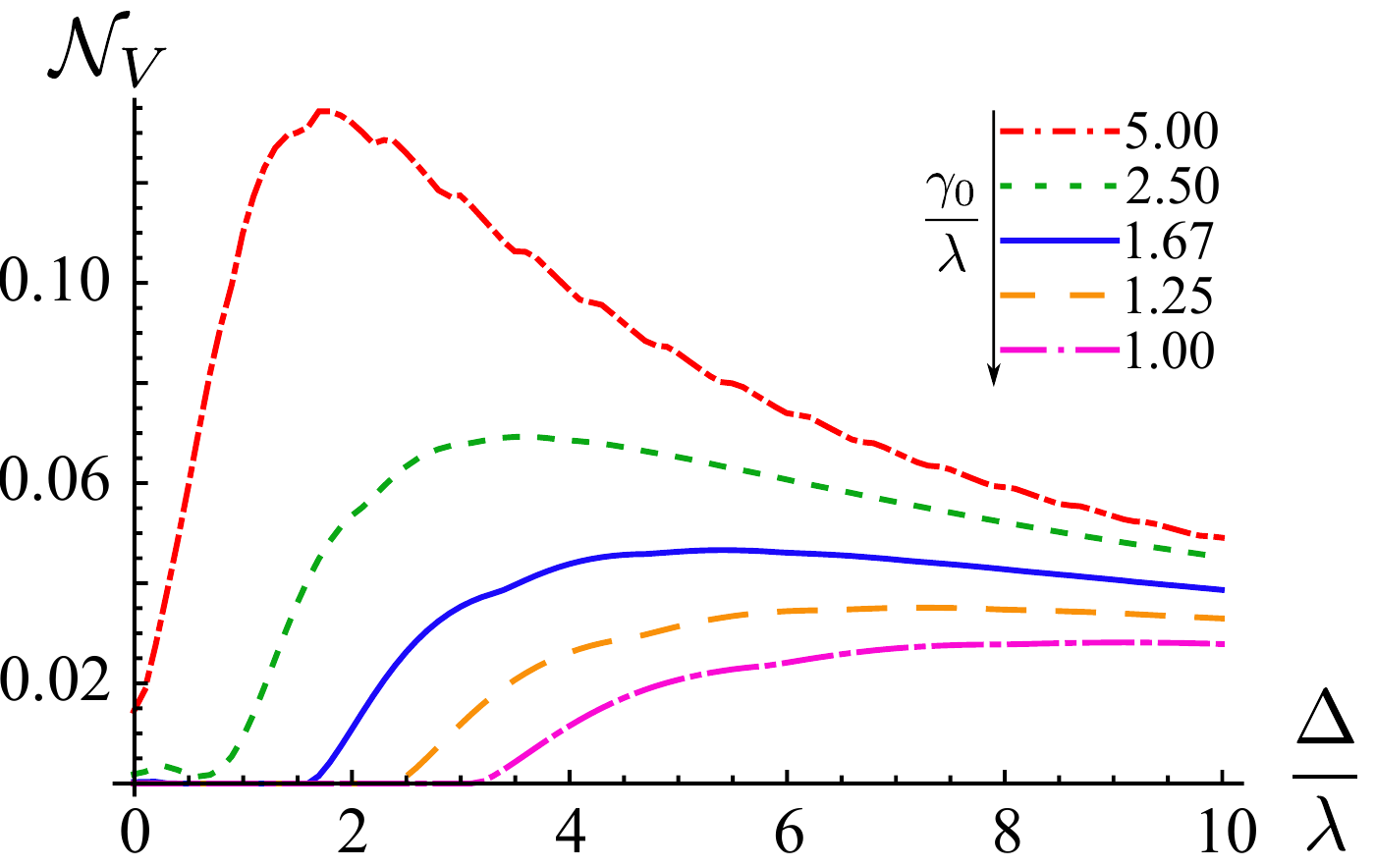}
\caption{(Color online) Panle {\bf (a)} Example of an ellipsoid
representing the image of a random map for a qubit. The canonical
basis of $\mathbb{R}^3$ is mapped onto the corresponding colored
circles highlighted on the ellipsoid, from which the volume can be
obtained. Panel {\bf (b)} shows the non-Markovianity measure $\mathcal{N}_V$ as a
function of the detuning, for the spontaneous emission dynamics
into a reservoir with Lorentzian spectral density (cf. {\it Example 1}).}
\label{trepunti}
\end{figure}

From a practical viewpoint, the experimental evaluation of ${\cal N}_V$ passes through the determination of the volume of the
set of evolved states  found from the contraction of $B_N$. This is an ellipsoid for the case
of a qubit.
As the number of evolved states that any realistic experimental implementation can sample is finite, it would be a precious piece of information to know
which are the best initial states to use in order to determine the set of physically accessible states at a given time $t$ of the dynamics
and its volume. For this purpose, let us consider $N$
initial Bloch vectors, evolved up to time $t$ and arranged as
the columns of a matrix $\mathbf{P}_t$. We have from Eq.~\eqref{affine_map}, that such vector evolves as $\mathbf{P}_t=\mathbf{A}_t\mathbf{P}_0+\mathbf{Q}_t$,
where the columns of $\mathbf{Q}_t$ are given by the
$\mathbf{q}_t$'s (which provides the form of the evolved state for a maximally
mixed initial condition). From this it follows that
$\deter{(\mathbf{P}_t-\mathbf{B}_t)(\mathbf{P}_t-\mathbf{B}_t)^T}=(\deter{\mathbf{A}_t})^2(\deter{\mathbf{P}_0})^2$. In turn, if we choose as initial Bloch vectors the elements of
any orthogonal basis in $\mathbb{R}^{N^2-1}$ plus the null vector
corresponding to the maximally mixed state, then their time
evolutions (arranged to form the matrix
$\mathbf{P}_t-\mathbf{Q}_t$) gives the determinant of the map,
from which the measure ${\cal N}_V$ easily follows. Therefore, the geometric measure of non-Markovianity
in Eq.~\eqref{NM_nostra} can be revealed experimentally by performing a
state tomography at different times for $N^2-1$ initial orthogonal
states. This will be sufficient to evaluate the change in volume of the accessible states without the need
for prior knowledge about the environment or the coupling. For the case of a qubit, this is illustrated in Fig.
\ref{trepunti} {\bf (a)}, where the three initial Bloch vectors
corresponding to the canonical basis of $\mathbb{R}^3$ are shown
to evolve into the extreme points on the semi-axis of the
ellipsoid that comprises all the possible accessible states of the evolution.
\noindent
{\it Systems with infinite dimensional Hilbert space}. We can
extend our idea to the less intuitive context of continuous
variable systems, in which the Hilbert space is infinite
dimensional and it is not possible to describe a state through a
finite number of parameters. However, the
restriction to Gaussian state and Gaussian-preserving processes
helps overcoming this issue.

We consider a system made of $n$ bosonic modes $k=1,..,n$, each described by the annihilation
and creation operators $\hat{a}_k$ and $\hat{a}^\dag_k$ [corresponding position and momentum operators
 $\hat{q}_k=1/\sqrt{2}(\hat{a}_k+\hat{a}^\dag_k)$ and
$\hat{p}_k=-i/\sqrt{2}(\hat{a}_k-\hat{a}^\dag_k)$, respectively].
Defining the vector of operators
$\mathbf{\hat{R}}=(\hat{q}_1,\hat{p}_1,\hdots,\hat{q}_n\hat{p}_n)^T$,
the commutation relations can be written as
$[\hat{R}_k,\hat{R}_l]=i \Omega_{kl}$ where $\Omega_{kl}$ are
elements of the symplectic matrix $\mathbf{\Omega}=\oplus_{k=1}^n
\boldsymbol{\omega}$ with
$\boldsymbol{\omega}=(\begin{smallmatrix}0 & 1 \\-1 & 0
\end{smallmatrix})$. A Gaussian state is completely characterized
by its first and second statistical moments, given respectively by
$\langle \mathbf{R}\rangle$ and the covariance matrix
$\boldsymbol{\sigma}$ defined as
\begin{equation}
\sigma_{kl}=\frac{1}{2}\langle\{\hat{R}_k,\hat{R}_l\}\rangle-\langle\hat{R}_k\rangle\langle\hat{R}_l\rangle.
\end{equation}\\
First moments can be adjusted to be null by local unitary operations. It can be shown that any evolution
resulting from the reduction of a symplectic evolution on a larger
Hilbert space can be described, in terms of second moments, by the equation
\begin{equation}
\sigma_t\rightarrow X_t^T\sigma_0 X_t + Y_t
\label{covariancemap}\end{equation} where $X$ and $Y$ are
$2n\times2n$ real matrices fulfilling the relation $Y+i\mathbf{\Omega}-i X^T \mathbf{\Omega}
X\geq 0$. Viceversa, any evolution of this kind may be interpreted
as the reduction of a larger symplectic evolution \cite{Wolf2005}.
In full analogy with what we did in \eqref{mapmatrix}, by choosing a basis $\lbrace G_j \rbrace$ for the space of $2n\times
2n$ matrices, we can write such map  as
\begin{equation}
\sigma_t =
\sum_{jk} {\rm Tr}[X_t^T G_k X_t G_j]{\rm Tr}[\sigma_0 G_k]G_j+ {\rm Tr}[Y_t G_j]G_j.
\label{vectorization}
\end{equation}
Eq.~\eqref{vectorization} can be recast as the $\mathbb{R}^{4N^2}$  affine
transformation $\mathbf{s}_0\rightarrow\mathbf{s}_t=\mathbf{X}_t\mathbf{s}_0+\mathbf{Y}_t$.
We then define a measure of non-Markovianity in a way fully analogous to what has been done above for the case of a discrete-variable system, {\it i.e.} as in Eq.~\eqref{NM_nostra} with th replacement ${\bf F}_t\to{\bf X}_t$.

If the evolution is unitary, the associated transformation has a
constant determinant, equal to one. 
For a single mode in a generic noisy Markovian channel, we have~\cite{Bibliopolis}
\begin{equation}
\sigma_t = e^{-\Gamma t}\sigma_0 + (1-e^{-\Gamma t})\sigma_{\infty},
\end{equation}
which gives $\deter{\mathbf{X}_t}=e^{-4\Gamma t}$. Therefore, every increase in $\deter{\mathbf{X}_t}$signals non-Markovianity.

Having introduced our formalism, we now illustrate our proposal with the aid of a few significant examples.

\noindent
{\it Example 1: Spontaneous emission into a leaky cavity}.
Consider a single two-level atom with transition frequency
$\omega_0$ interacting with a vacuum electromagnetic field having
a Lorentzian spectral density (mimicking a leaky cavity) \cite{Breuer2010}. Taking
\begin{equation}
J(\omega)=\frac{1}{2\pi}\frac{\gamma_0\lambda^2}{(\omega_0-\Delta-\omega)^2+\lambda^2},
\label{spectraldens}\end{equation} where $\Delta$ is the detuning
between the atomic and the cavity frequency, the atomic state
at time $t$ reads
\begin{equation}
\hat{\rho}_A(t)=\left(\begin{matrix}
|\Gamma(t)|^2\rho_0^{++} & \Gamma(t)\rho_0^{+-}\\
\Gamma(t)^*\rho_0^{-+} & (1-|\Gamma(t)|^2)\rho_0^{++}+\rho_0^{--}
\end{matrix}\right)
\label{rhotjaynes}\end{equation}
with $\Omega_\pm=\Delta -i\lambda\pm \sqrt{(\Delta -i \lambda )^2+2 \gamma _0 \lambda }$ and 
\begin{equation}
\Gamma(t)=\frac{e^{-\frac{i t\Omega_-}{2}}\Omega_+ -
e^{-\frac{i t \Omega_+}{2}}\Omega_-}{2(\Omega_+-\Delta+i\lambda)}.
\end{equation}
The evolution of the Bloch vector is ruled by
\begin{equation}
\mathbf{A}_t=
\begin{pmatrix}
\mathfrak{Re}\Gamma(t) & \mathfrak{Im}\Gamma(t) & 0 \\
-\mathfrak{Im}\Gamma(t) & \mathfrak{Re}\Gamma(t) & 0 \\
0 & 0 & |\Gamma(t)|^2
\end{pmatrix},\label{map_jaynes}
\end{equation}
whose determinant, $|\mathbf{A}_t|=|\Gamma(t)|^4$, is shown in
Fig.~\ref{mnm} against the dimensionless time $\lambda t$. The corresponding non-Markovianity measure ${\cal
N}_V$ is reported in Fig.~\ref{trepunti}, from which it is clear
that a strongly non-Markovian behavior is found for a resonant
coupling and invoking the so-called good-cavity limit $\gamma_0 \gg
\lambda$. A similar result is obtained with the RHP measure,
\cite{Huelga2010} which is given by the integral of $(1/2)\mathfrak{Re}[\partial_t\ln\Gamma(t)]$. This is in agreement
with BLP as well, which turns out to depend on $|\Gamma(t)|$~\cite{Breuer2010}. 

\noindent
{\it Example 2: Pure dephasing}. Let us consider a qubit
undergoing a purely dephasing dynamics, expressed in terms of  a
decoherence factor $\nu(t)$ as
\begin{equation}
\phi^{(d)}_{t}[\rho(0)]=\left(\begin{matrix}
\rho^{++} & \nu(t)\rho^{+-}\\
\nu(t)\rho^{-+} & \rho^{--}
\end{matrix}\right).
\label{mapdephasing}
\end{equation}
In this case, the BLP and RHP measures coincide~\cite{cinesi,Plastina_Goold}. In fact, the trace distance optimised over the initial qubit preparation
gives us $D[\rho_1^{opt},\rho_2^{opt}]=|\nu(t)|$. In turn, $|\nu(t)|$ is exactly the value of the concurrence between a system qubit and ancilla, initially prepared
in a maximally entangled state and undergoing a unilateral dephasing mechanism. As for our proposal,  the evolution of the Bloch vector is determined by the matrix
\begin{equation}
\mathbf{A}_t=\left(\begin{matrix}
\mathfrak{Re} \nu(t) & \mathfrak{Im} \nu(t) & 0 \\
- \mathfrak{Im} \nu(t) & \mathfrak{Re} \nu(t) & 0 \\
0 & 0 & 1
\end{matrix}\right) \, ,
\end{equation}
which has $\left|\deter{\mathbf{A}_t}\right|=|\nu(t)|^2$. The
geometric measure ${\cal N}_V$ thus gives the same behavior predicted by the
other two quantifiers.

We have proposed a geometrically motivated quantifiers of non-Markovianity that is explicitly linked to the variations in the volume of the physical states dynamically accessible by a given open open system. From an information theoretical perspective, such measure, which provides predictions that are, in general, in qualitative agreement with those coming from some of the most popular tools for the characterisation of non-Markovianity proposed to date, is linked to the loss/regain of classical information oven the evolving system. We have shown how an estimate of the proposed measure is possible through only a polynomial number of measures, while the proposed formal quantifier itself enjoys a straightforward extension to the Gaussian continuous-variable scenario. We have illustrated our proposal through a series of examples. We hope that the appealing aspects of practicality and intuitive nature of our proposal will soon spur the attention of the community interested in open-system dynamics.

SL thanks the Centre for Theoretical Atomic, Molecular and Optical Physics for hospitality during the early stages of this work. MP thanks the UK EPSRC for a Career Acceleration Fellowship and a grant awarded under the ``New Directions for Research Leaders" initiative (EP/G004579/1).

\end{document}